\documentclass[a4paper, 10pt, sfdefaults=false]{scrartcl}

\usepackage{fullpage}

\usepackage{lmodern}
\usepackage[british]{babel}
\usepackage[affil-it]{authblk}

\usepackage{amsmath}
\usepackage{mathtools}
\usepackage{amssymb}

\newcommand{\Lap}{\Delta}
\newcommand{\Dim}{D}
\newcommand{\spherearea}[1]{\ensuremath{S_{#1}}}

\newcommand{\reals}{\ensuremath{\mathbb{R}}}

\DeclarePairedDelimiterX\abs[1]\lvert\rvert{
  #1
}\DeclarePairedDelimiterX\norm[1]\lVert\rVert{
  #1
}

\makeatletter
\def\blfootnote{\gdef\@thefnmark{}\@footnotetext}
\makeatother

\usepackage[backend=biber,
url=false,
doi=false,
style=ext-numeric,
articlein=false,
]{biblatex}
\usepackage{hyperref}

\title{Some exact solutions of the Schrödinger--Poisson system in spaces of constant sectional curvature}
\author{Richard Chapling}
\affil{Department of Applied Mathematics and Theoretical Physics, \\ University of Cambridge, Cambridge, England
}

\begin{document}


\maketitle

\blfootnote{\textbf{2020 Mathematics Subject Classification:} 35C05 (Primary) 35C08, 35G50, 35J10, 35Q40, 35R01, 58J05 (Secondary)}

\blfootnote{\textbf{Keywords:} Schrödinger--Newton system, Schrödinger--Coulomb system, Explicit solution, Closed form solution}

\begin{abstract}
  We consider the Schrödinger--Poisson system on the complete, simply-connected Riemannian manifolds of constant sectional curvature.
  We obtain closed-form stationary spherically-symmetric solutions for the homogeneous equations for certain dimensions, and give some basic examples of solutions with a nonzero background.
\end{abstract}

\section{Introduction}
\label{sec:introduction}

The system of equations
  \begin{align*}
    i\dot{\psi} &= -\Lap \psi + \alpha V \psi \\
    -\Lap V &= \abs{\psi}^{2}
  \end{align*}
is often referred to as the \emph{Schrödinger--Newton system} when \(\alpha<0\) (the self-attractive case), and the \emph{Schrödinger--Coulomb} system when \( \alpha > 0 \) (the self-repulsive case).
Since the second equation is Poisson's equation, we refer to the general case with \(\alpha \neq 0\) as the \emph{Schrödinger--Poisson system}.

This system in both of these forms has been studied from various points of view over the years in both time-dependent and time-independent forms, for the Schrödinger--Newton version \cite{Penrose:1996GravQStateRed}, \cite{MorozPenroseTod:1998SphSymSN}, \cite{Choquard:2007fk}, the Schrödinger--Coulomb version \cite{BenciFortunato:1998EvalSchMax}, \cite{CocliteGeorgiev:2003MaxSchr}, or with both signs as Schrödinger--Poisson \cite{CingolaniWeth:2016PlanarSP}. Integrating the second equation leads to a nonlocal nonlinear equation, the Hartree equation, which has also been also well-studied \cite{FrohlichLenzmann:2004MeanFieldBoseHartree}.

However, since these equations are nonlinear, exact solutions are difficult to come by.
In this note we will obtain exact solutions to the system when \( \psi(t,x) = e^{-i\omega t} u(x) \) is stationary, and \(u\) is real and spherically symmetric, i.e.
\begin{subequations}
  \label{eq:schrodinger-poisson-equations+background}
  \begin{align}
    -\Lap u + \alpha V u &= \omega u \\
    \label{eq:schrodinger-poisson-equations+background-poisson}
    -\Lap V &= u^{2} ;
  \end{align}
\end{subequations}
This is a natural simplification to make for static solutions since the probablity current depends on the phase of the wavefunction, so fixed phase gives no current.

It was demonstrated by Choquard, Stubbe and Vuffray~\cite{ChoquardStubbeVuffray:2008SNODE} that for \( \alpha < 0 \), \eqref{eq:schrodinger-poisson-equations+background} admits exact solutions in \(6\)-dimensional Euclidean space, which take the simple form
\begin{equation}
  \label{eq:CSV-solution}
  \psi(x) = A(1+\abs{x}^{2})^{-2}
\end{equation}
for a particular constant \(A\).
We observe that solutions of this type are spherically symmetric, smooth, and square-integrable.
  Further, we may generate other solutions by rescaling with appropriate adjustment of \(A\).
  We will obtain similar solutions in hyperbolic space, in particular when \( \Dim = 3 \).

Some exact solutions to the Schrödinger--Newton equations in low dimensional Euclidean space were also given by Bougoffa \textit{et al.}~\cite{BougoffaKhanferBougouffa:2023gz}, although these are too singular to be square-integrable.

We attempt to solve \eqref{eq:schrodinger-poisson-equations+background} in the form
\begin{align*}
  \alpha V - \omega &= \frac{\Lap u}{u} \\
  -\Lap V - u^{2} &= 0 ,
\end{align*}
which allows us to determine \(V\) from \(u\), and use the second equation as a consistency condition.\footnote{This substitution for \(V\) is equivalent to Bougoffa \textit{et al.}~\cite{BougoffaKhanferBougouffa:2023gz}'s equations (2.8) and (3.4), to which they give particular solutions.}
We use this procedure first to recover \eqref{eq:CSV-solution}, then to find a solution in hyperbolic space.
Both of these require \(\alpha<0\), so in the last section we consider solutions to the equations with a background potential, which allows us to produce exact solutions for the \( \alpha>0 \) case as well.

It is worth noting that the value of \(\omega\) is not well-defined without further condition on \(V\), so in the following we will fix \(\omega\) as the limit as \( r \to \infty\) of \( -\Lap u / u \), when it exists.

We write \( \spherearea{\Dim-1} \) for the area of the unit sphere in \(\Dim\) dimensions.
The total mass of a solution is conventionally given by \( N[u] \coloneq \int  u^{2} \).

\section{Flat space}
\label{sec:flat-space}

\newcommand{\ech}{c}

Starting from \eqref{eq:CSV-solution}, we try the more general \( u(x) = A(1+\abs{x}^{2})^{n/2} \) in \(\Dim\) dimensions.
We will abbreviate \( \ech = (1+r^{2})^{1/2} \). 
We compute
\begin{equation*}
  \alpha V-\omega = \frac{\Lap u}{u} = - \frac{n(n-2)}{\ech^{4}} + \frac{n(\Dim+n-2)}{\ech^{2}} ,
\end{equation*}
so by our convention \( \omega = 0 \).
Then,
\begin{equation*}
  \alpha \Lap V = \frac{24n(n-2)}{\ech^{8}} + \frac{4n(\Dim n - 8n - 4\Dim + 16)}{\ech^{6}} - \frac{2n(\Dim-4)(\Dim+n-2)}{\ech^{4}} .
\end{equation*}
Inserting this into \eqref{eq:schrodinger-poisson-equations+background-poisson} and clearing denominators, we find we need
\begin{equation*}
  24n(n-2) + 4n(\Dim n - 8n - 4\Dim + 16) \ech^{2}  - 2n(\Dim-4)(\Dim+n-2) \ech^{4} +  \alpha A^{2} \ech^{2n+8} = 0 .
\end{equation*}

For this equation to be satisfied for every \(\ech\), the last term must cancel one of the first three terms, and the other two must vanish.
\(n=0\) and \(n=1\) both lead us being unable to cancel the remaining terms, so the first term cannot vanish.
Therefore the only way to cancel the first term is to have \( n = -2 \), and \( A = 24(-\alpha)^{-1/2} \).
For both of the other terms to vanish, we see that the only way is if \( \Dim = 6 \), as expected.
Hence the only exact solutions of this form are
\begin{align*}
  \label{eq:schrodinger-newton-flat-exact-solution}
  u_{CSV}(x) &= \frac{24(-\alpha)^{-1/2}}{\left( 1 + \abs{x}^{2} \right)^{2}} & (\Dim = 6) .
\end{align*}
The total mass of this solution is
\begin{equation*}
  N[u] = \int_{\reals^{6}} (u(x))^{2} \, dx = \spherearea{5} \frac{96}{-\alpha}
\end{equation*}
and the frequency \(\omega\) vanishes. As noted above, the rescaling properties of the equations give us that
\begin{equation*}
  u_{a}(x) \coloneqq a^{-2} u_{CSV}(x/a)
\end{equation*}
is also a solution with \( \omega = 0\), and we can check that the masses are related by
\begin{equation*}
  N[u_{a}] = a^{2} N[u] .
\end{equation*}

Hence we have re-obtained the result of Choquard, Stubbe and Vuffray~\cite{ChoquardStubbeVuffray:2008SNODE}.

We can use the same technique to produce singular solutions: if instead we try \( u = A r^{n} \), we obtain the solution \( u_{s}(r) = 2(\Dim-4) (-\alpha)^{-1} r^{-2} \), which is singular at the origin and has infinite total mass, but is available in any \( \Dim \neq 4 \).
The frequency of this solution is also \( \omega = 0 \).

\section{Hyperbolic space}
\label{sec:exact-hyperbolic-space}

Again we assume we are in the Newtonian case \( \alpha < 0 \).
We use the notation
\begin{align*}
  S(r) &= (-\kappa)^{-1/2} \sinh( (-\kappa)^{1/2} r) \\
  C(r) &= \cosh( (-\kappa)^{1/2} r) \\
  T(r) &= S(r) / C(r) =  (-\kappa)^{-1/2} \tanh(  (-\kappa)^{1/2} r) .
\end{align*}
The significance of these functions is that the metric of hyperbolic space in spherical hyperboloid coordinates is
\begin{equation*}
  ds^{2} = dr^{2} + S(r)^{2} \, d\Sigma_{\Dim-1}^{2} ,
\end{equation*}
where \( d\Sigma_{\Dim-1}^{2} \) is the area measure on the \((\Dim-1)\)-dimensional hypersphere, and then the scalar Laplacian is
\begin{align*}
  \Lap \phi
  &= \frac{1}{(S(r))^{\Dim-1}} \frac{\partial}{\partial r} \left( S(r)^{\Dim-1} \frac{\partial \phi}{\partial r} \right) + \frac{1}{(S(r))^{2}} \Lap_{S_{\Dim-1}} \phi \\
  &= \frac{\partial^{2} \phi}{\partial^{2} r} + \frac{\Dim-1}{T(r)} \frac{\partial \phi}{\partial r} + \frac{1}{(S(r))^{2}} \Lap_{S_{\Dim-1}} \phi .
\end{align*}
See, e.g. the discussion in the paper \cite{Cohl:2012ly}, §~2.

Then we find if \( u = A(C(r))^{n} \), we obtain
\begin{equation*}
  \alpha V - \omega = (-\kappa) n (\Dim+n-1) - \frac{(-\kappa)n(n-1)}{C^{2}} .
\end{equation*}
We see that unlike the Euclidean case, we have a nonzero \(\omega\),
\begin{equation*}
  \omega = -(-\kappa) n (\Dim+n-1) .
\end{equation*}

Differentiating again we obtain
\begin{equation*}
  \alpha \Delta V = \frac{ 6n(n-1) (-\kappa)^{2} }{C^{4}} + \frac{ 2n(n-1)(\Dim-3) (-\kappa)^{2} }{C^{2}} .
\end{equation*}
We therefore conclude by a similar argument to the flat case that when \( \Dim = 3 \), if we choose \( n=-2 \) we obtain an exact solution to the Schrödinger--Newton system,
\begin{equation}
  \label{eq:schrodinger-newton-hyperbolic-exact-solution}
  u_{1}(x) = \frac{6(-\kappa)(-\alpha)^{-1/2}}{(C(\abs{x}))^{2}} .
\end{equation}
This has total mass
\begin{equation*}
  N[u_{1}] = \spherearea{2} \frac{36(-\kappa)^{2}}{-\alpha} \int_{0}^{\infty} \frac{(S(r))^{2}}{(C(r))^{4}} \, dr
  = \spherearea{2} \frac{12 (-\kappa)^{1/2}}{-\alpha} .
\end{equation*}

We observe the following properties of this solution:
\begin{enumerate}
\item
  There is no way to scale \(r\) and \(u\) to produce a new solution: for each curvature \(\kappa\) and coupling \(\alpha\) we obtain a single solution.
  This is essentially a consequence of the hyperbolic Laplacian not being scale-invariant, as one would expect since hyperbolic space itself is not scale-invariant.

\item
  Since \( u_{1} \propto -\kappa \), as \( \kappa \to 0 \) it shrinks uniformly to \(0\): we do not get a useful Euclidean limit.

\item
  On the other hand, unlike the solutions given by Bougoffa \textit{et al.} \cite{BougoffaKhanferBougouffa:2023gz} to the one- and two-dimensional equations, it is square-integrable.
\end{enumerate}

Another \(u\) we might try is \( A(S(r))^{n} \).
The condition for this to be a solution is
\begin{equation*}
  \alpha A^{2} S^{2n+4} - 2 (-\kappa)^{2} n (\Dim+n-2) ( \Dim-4 + (\Dim-3) S^{2} ) = 0 ,
\end{equation*}
so now there are two options:
\begin{itemize}
\item
  \( n=-2 \), \( \Dim=3 \), \( A = 2(-\kappa)(-\alpha)^{-1/2} \)
  \begin{equation*}
    u_{2}(x) \coloneq \frac{2(-\kappa)(-\alpha)^{-1/2}}{(S(\abs{x}))^{2}} ,
  \end{equation*}
\item
  \( n=-1 \), \( \Dim=4 \), \( A = 2^{1/2}(-\kappa)(-\alpha)^{-1/2} \)
  \begin{equation*}
    u_{3}(x) \coloneq \frac{2^{1/2}(-\kappa)(-\alpha)^{-1/2}}{S(\abs{x})} .
  \end{equation*}
\end{itemize}

Both of these functions are smooth except at \(0\), and both have divergent integrals, \(u_{2}\) from small \(r\) and \(u_{3}\) from large \(r\).

Therefore we have the following three solutions to the attractive case:
\begin{align*}
  u_{1}(x) &= \frac{6(-\kappa)(-\alpha)^{-1/2}}{(C(\abs{x}))^{2}} && (\Dim = 3) \\
  u_{2}(x) &= \frac{2(-\kappa)(-\alpha)^{-1/2}}{(S(\abs{x}))^{2}} && (\Dim = 3) \\
  u_{3}(x) &= \frac{2^{1/2}(-\kappa)(-\alpha)^{-1/2}}{S(\abs{x})} && (\Dim = 4) .
\end{align*}

\subsection{Hypersphere}
\label{sec:hypersphere}

The calculations are essentially unchanged for a manifold with constant sectional curvature \(\kappa>0\), i.e. a hypersphere: the operators are of the same form, but
\begin{equation*}
  S(r) = \frac{1}{\kappa^{1/2}} \sin ( \kappa^{1/2} r )
\end{equation*}
etc.

It follows that the hyperbolic results still work for hyperspheres, but \(1/C(r)\) diverges when \\
\( r = \pi/(2\kappa^{1/2}) \), so  is singular on the equator, while \(1/S(r)\) diverges at the two antipodes \( r = 0 \) and \( r = \pi/\kappa^{1/2} \).
Now, however, \(u_{3}\) has a finite integral, namely \( 4\kappa/(-\alpha) \).

At least one singularity is inevitable for this system on a compact manifold: if we suppose that \(u\) is continuous, elliptic regularity implies that \(V\) is \(C^{2}\) and hence we can integrate \eqref{eq:schrodinger-poisson-equations+background-poisson} over the whole manifold:
\begin{equation*}
  \int u^{2} = - \int \Lap V = 0
\end{equation*}
by Stokes's theorem, a contradiction.

\section{With background}
\label{sec:with-background}

In the self-repulsive case, there can be no bound states in \( H^{1} \).
In the flat case with \(\Dim > 2\), this is straightforward to show using Pohozaev identities analogously to Georgiev and Venkov~\cite{georgiev2010symmetryuniquenessminimizershartree} §~4: we obtain the identities
\begin{align*}
  T - \omega N + \alpha Q &= 0 \\
  (\Dim-2)T -\Dim \omega N + \tfrac{1}{2} (\Dim+2) \alpha Q &= 0 ,
\end{align*}
where \( T \coloneq \int \abs{\nabla u}^{2} \), \( N \coloneq \int u^{2} \), \( Q \coloneq \int u^{2} (-\Lap)^{-1} u^{2} \).
From here we find
\begin{equation*}
  4T + (\Dim-2)\alpha Q = 0
\end{equation*}
Since \(T>0\) and \(Q>0\) for \( \Dim > 2\), we must have \(\alpha < 0 \) to have a bound state of this type.

Therefore for solutions to the repulsive case, we allow for a background potential in Poisson's equation, replacing \eqref{eq:schrodinger-poisson-equations+background-poisson} by
\begin{equation*}
  -\Lap V = u^{2} + \rho .
\end{equation*}
Obviously if we allow arbitrary \(\rho\) essentially any trial \(u\) will work, so we look only for solutions with the simplest \(\rho\) and \(u\), without singularities.

\subsection{Flat space}
\label{sec:flat-space-bg}

Looking again at the first calculation, we have three terms, one which we can cancel with \(\alpha u^{2}\), the other by choosing \(n\) or \(\Dim\), leaving one to become \(\rho\).

Again \(n=0\) and \(n=2\) leave us with more than we want, so we choose \(n\) to eliminate one of the other terms.

For \(n=-2\), there is no integer \(D\) that cancels the \( c^{-6} \) term.

For \(n=-3\), we obtain solutions to the repulsive system in \( \Dim=4 \) and \( \Dim = 5 \), namely
\begin{align*}
  u &= \frac{12\alpha^{-1/2}}{c^{3}} , & \rho &= -\frac{360\alpha^{-1}}{c^{8}} && (\Dim = 4) \\
  \intertext{and}
    u &= \frac{60^{1/2}\alpha^{-1/2}}{c^{3}} , & \rho &= -\frac{360\alpha^{-1}}{c^{8}} && (\Dim = 5) .
\end{align*}
This solution does not have finite mass.

For \(n=-4\), we know that \(\Dim=6\) gives \(\rho=0\), but we can choose \(\Dim=4\) instead, and find a solution to the attractive system:
\begin{align*}
  u &= \frac{24(-\alpha)^{-1/2}}{c^{4}} , & \rho &= \frac{256(-\alpha)^{-1}}{c^{6}} && (\Dim = 4) .
\end{align*}
This solution also does not have finite mass.

\subsection{Hyperbolic space}
\label{sec:hyperbolic-space-bg}

We consider the solutions containing \(C\).
There are only three terms here:
\begin{equation*}
  -\rho = A^{2}C^{2n} + (-\kappa)^{2}\alpha^{-1} \frac{6n(n-1)}{C^{4}} + (-\kappa)^{2}\alpha^{-1} \frac{2n(n-1)(\Dim-3)}{C^{2}} .
\end{equation*}

Choosing \(n=-2\) gives the solution to the attractive equations we found earlier,
\begin{align*}
  u &= \frac{6(-\kappa)(-\alpha)^{-1/2}}{C^{2}} & \rho &= \frac{12(\Dim-3)(-\kappa)^{2}(-\alpha)^{-1}}{C^{2}} .
\end{align*}
Since in gravitation mass is always nonnegative, this \(\rho\) makes sense for \( \Dim \geq 3 \), but only has finite mass if \( \Dim \leq 4 \).

Choosing \(n=-1\) gives
\begin{align*}
  u &= 2(-\kappa) \left( \frac{\Dim-3}{-\alpha} \right)^{1/2} \frac{1}{C} & \rho &= \frac{12(-\kappa)^{2} (-\alpha)^{-1} }{C^{4}} .
\end{align*}
The situation here is more complicated:
\begin{itemize}
\item
  If \( \Dim < 3 \), for \(u\) to be real we need \( \alpha > 0 \), so \(\rho\) is negative.
  This is fine: we can interpret it as a negative charge.

\item
  If \( \Dim = 3 \), \(u=0\) so the algorithm does not work.

\item
  If \( \Dim > 3 \), \(u\) is real when \( \alpha < 0 \), and \(\rho\) is positive.
\end{itemize}

In this case, \(u\) only has finite mass for \(\Dim<3\).

\paragraph{Another Euclidean solution}
If \(\Dim=1\), the space is the same as \( \reals \): putting \(\Dim=1\) and \(\kappa = -1/R^{2} \) (representing the radius of curvature), we have
\begin{align*}
  u &= \frac{8^{1/2} \alpha^{-1/2} }{R^{2}\cosh(r/R)} & \rho &= -\frac{12 \alpha^{-1} }{(R\cosh(r/R))^{4}} .
\end{align*}
for the repulsive case. As with the Euclidean solutions \eqref{eq:CSV-solution}, this is a family related by scaling. The mass is
\( N[u] = 16/(R^{3}\alpha) \), the same as the integral of \(-\rho\).

\subsection{Hypersphere}
\label{sec:hypersphere-bg}

None of our solutions are regular everywhere, there is a trivial solution \( u^{2} = -\rho = \text{const.} \)

\section{Conclusion}
\label{sec:conclusion}

We have obtained several simple solutions to the Schrödinger--Poisson system in various dimensions.
Each solution requires \(\alpha\) to have a particular sign: it is valid for either the attractive or repulsive case.

The \(6\)-dimensional Euclidean solution is derivable from transforming the equations into an autonymous system using the scaling symmetry of the equations and constructing a Hamiltonian (\cite{ChoquardStubbeVuffray:2008SNODE}, Appendix~A).
This is not possible for the \(3\)-dimensional hyperbolic solution since the radial equations possess no Lie point symmetries, so it would be interesting to have a corresponding interpretation.

\paragraph{Acknowledgements}
The author would like to thank J. F. Wang for stylistic suggestions and David Stuart for help with presentation.

This work has been partially supported by STFC consolidated grant ST/X000664/1.

\printbibliography

\end{document}